\documentclass[prl,twocolumn,notitlepage,tightenlines]{revtex4}
\usepackage{amsmath}
\usepackage{amssymb,amsfonts}
\usepackage{bm}
\usepackage[mathcal]{euscript}
\usepackage{graphicx}
\usepackage{subfigure}
\usepackage{psfrag}
\newcommand{\comment}[1]{}

\newcommand{\etal}{\textit{et~al.}}

\newcommand{\mathnotation}[2]{\newcommand{#1}{\ensuremath{#2}}}

\newcommand{\goodgap}{%
	\hspace{\subfigtopskip}%
	\hspace{\subfigbottomskip}}

\mathnotation{\pd}{\partial}			% Partial derivative
\mathnotation{\ee}{{\mathrm e}}			% e
\mathnotation{\imi}{\mathrm{i}}			% i
\mathnotation{\ldef}{\mathrel{\raisebox{.069ex}{:}\!\!=}}% Left define
\mathnotation{\rdef}{\mathrel{=\!\!\raisebox{.069ex}{:}}}% Right define
\mathnotation{\dint}{\,{\mathrm{d}}}		% Differential, in an integral

\mathnotation{\grad}{\nabla}			% Nabla (grad) symbol
\mathnotation{\curl}{\grad\times}		% The curl
\mathnotation{\lapl}{\nabla^2}			% Laplacian symbol

\renewcommand{\time}{t}				% Time
\mathnotation{\ssigma}{\Sigma}			% Product of braid matrices
\mathnotation{\ip}{i}				% Counter for braid elements
\mathnotation{\rc}{r}				% Position vector comp
\mathnotation{\rv}{\bm{\rc}}			% Position vector
\mathnotation{\nn}{n}				% Number of particles

\begin{document}

\title{Measuring Topological Chaos}
\author{Jean-Luc Thiffeault}
\email{jeanluc@imperial.ac.uk}
\affiliation{Department of Mathematics, Imperial College London,
SW7 2AZ, United Kingdom}

%\keywords{chaotic mixing, topological chaos}
%\pacs{47.52.+j, 05.45.-a}

\begin{abstract}
The orbits of fluid particles in two dimensions effectively act as topological
obstacles to material lines.  A spacetime plot of the orbits of such particles
can be regarded as a braid whose properties reflect the underlying dynamics.
For a chaotic flow, the braid generated by the motion of three or more fluid
particles is computed.  A ``braiding exponent'' is then defined to
characterize the complexity of the braid.  This exponent is proportional to
the usual Lyapunov exponent of the flow, associated with separation of nearby
trajectories.  Measuring chaos in this manner has several advantages,
especially from the experimental viewpoint, since neither nearby trajectories
nor derivatives of the velocity field are needed.
\end{abstract}

\maketitle

\textcite{Boyland2000} showed that strong chaos can be induced through
topological obstacles moving in a two-dimensional fluid.  Their system
consisted of three rods whose repeated interchange led to complicated braiding
of material lines.  Chaos ensued as a consequence of the topology:
Thurston--Nielsen theory~\cite{Thurston1988,Boyland1994} guarantees the
existence of a region of the flow which has pseudo-Anosov dynamics.  In
practical terms this means this region has extremely strong chaotic
properties---almost every point exhibits exponential stretching---leading to
good mixing~\cite{Aref1984}.  Thurston--Nielsen theory does not say how large
the pseudo-Anosov region is (it could even have zero measure), but in physical
and numerical experiments it has usually been found to be sizeable and
localized near the rods (though see~\cite{MattFinn2003b} for an example where
the region is too small to be important for mixing).

Physically speaking, material lines ``snag'' on the rods, and if the rods have
a complicated braiding motion then the length of material lines must grow
exponentially.

More recently, \textcite{Boyland2003} applied the theory where vortices serve
as topological obstacles. Several vortices orbited each other and the authors
classified the braiding properties of the different configurations.
\textcite{Vikhansky2003} pushed this a step further: he studied the properties
of freely-moving rods (which he called discs) in a two-dimensional cavity
flow.  In this case the braiding arises from the chaotic motion of the rods.
By mapping each element of the braid group to its matrix representation, a
Lyapunov exponent can be found which characterizes the vigor of topological
chaos.

In this letter we apply topological techniques as diagnostic tools to quantify
chaos in general two-dimensional dynamical systems.  In the absence of
diffusion, in a two-dimensional bounded domain any fluid particle (Lagrangian
tracer) is a topological obstacle to material lines.  That is, if we choose a
material line connected to the boundary and a reference fluid particle (not on
the material line), both moving with the fluid, then the material line must
bend around the fluid particle as they move.  Deterministic motion forbids the
crossing of the line and the fluid particle, since at the moment of crossing
the fluid particle would have to belong to the material line, and it must thus
have belonged to it for its entire history.

Since they are topological obstacles, any $\nn$ fluid particles can be seen as
candidates for topological chaos: the motion of the particles can exhibit
complex braiding.  Thus, we can look for the presence of topological chaos by
following $\nn$-tuplets of fluid particles and recording their braiding
history---an ordered sequence of braid group generators.  We then use an
appropriate matrix representation of the braid group to express this sequence
as a matrix product.

From random matrix theory, we can expect the resulting sequence to have
eigenvalues that grow or decay exponentially with well-defined Lyapunov
exponents, which we call braiding exponents.  If one of these exponents is
positive, then we say that the system exhibits topological chaos.  We will
show by an example that the magnitude of this largest exponent is proportional
to the usual Lyapunov exponent of the flow, defined in terms of separation of
nearby trajectories.  The work presented here is meant to be a step in the
development of what was aptly described as ``topological kinematics''
by~\textcite{Boyland2000}.

An obvious advantage of this method over measuring the chaotic properties of a
system by computing the usual Lyapunov exponents is that there is no need for
closely-spaced trajectories.  If an experiment yields data for a large number
of fluid trajectories, then all possible $\nn$-tuplets can be used to compute
the topological chaos properties (as long as they belong to the same chaotic
region).  Neither the velocity field nor its spatial derivatives are needed;
the only issue is whether the time series is of sufficient length.  (For
convenience we use the language of two-dimensional fluids, but the method
presented here can be applied to any two-dimensional deterministic flow.)

Historically, the focus was on periodic particle orbits before moving rods
were considered~\cite{Boyland1994}.  Here the new ingredient is that the
orbits are chaotic and determined explicitly by numerical methods.  The
association of arbitrary particle orbits with braids group elements was first
proposed by~\textcite{Gambaudo1999}.

We now describe the method, which we mean to keep as simple as possible for
easy implementation.  This is essentially the same technique as used
by~\textcite{Boyland2003} for the braiding of vortices
and~\textcite{Vikhansky2003} for freely-moving discs.  Our goal is to map the
motion of~$\nn$ particles onto elements of the braid group.

First project the position of the particles onto any fixed \emph{reference}
line (which we choose to be the horizontal axis), and label the particles
by~$\ip=1,2,\ldots,\nn$ in increasing order of their projection~%
\footnote{Unlike~\cite{Boyland2003}, we are not dealing with periodic
orbits. so a change of reference line does not simply lead to conjugation by
another braid word.  However, the braid word is changed by only a finite
number of braid generators, which does not influence the braiding exponent.}.
A crossing occurs whenever two particles interchange position on the reference
line.  A crossing can occur as an ``over'' or ``under'' braid, which for us
means a clockwise or counterclockwise interchange.  We define the
braid~$\sigma_\ip$ as the clockwise interchange of the $\ip$th and $(\ip+1)$th
particles, and~$\sigma_\ip^{-1}$ as their counterclockwise interchange,
for~$\ip=1,\ldots,\nn-1$.  These elementary braids are the generators of the
Artin~$\nn$-braid group~\cite{Murasugi}.

\begin{figure}
\psfrag{sii}{$\sigma_\ip^{-1}$}
\psfrag{si}{$\sigma_\ip$}
\psfrag{i}{$\ip$}
\psfrag{i+1}{$\ip+1$}
\psfrag{i+2}{$\ip+2$}
\subfigure[]{%
\includegraphics[width=.18\textwidth]{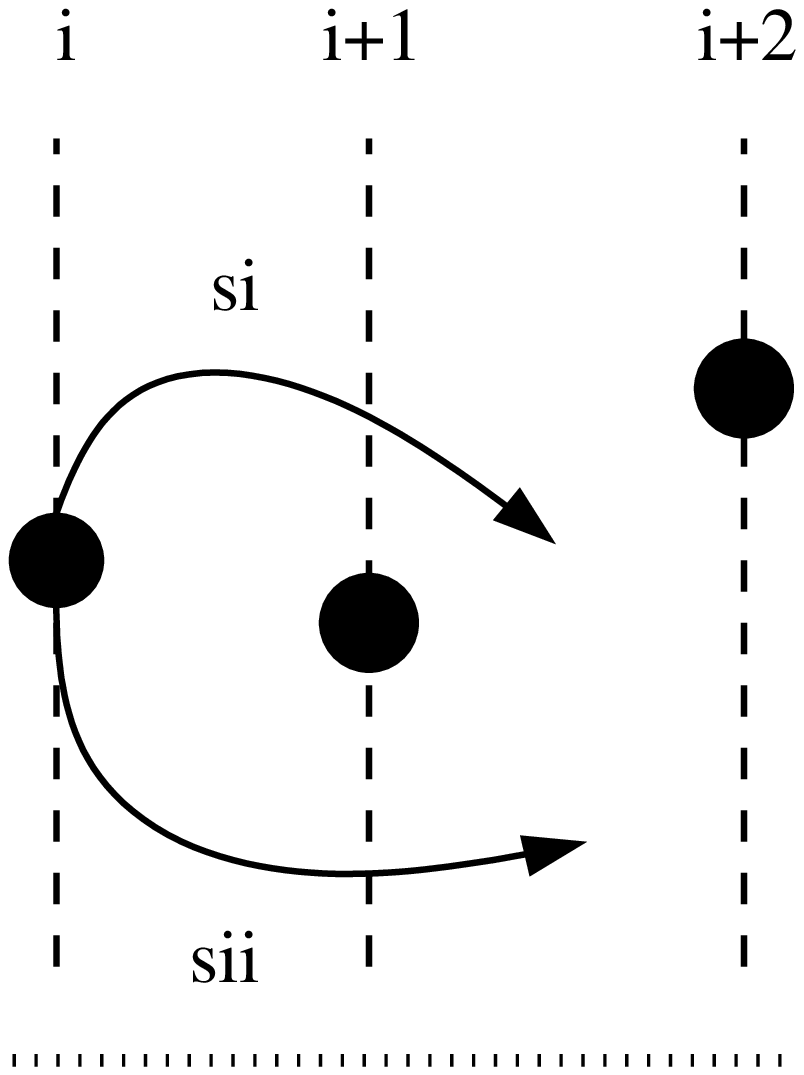}
\label{fig:crossinga}
}\goodgap\goodgap%
\subfigure[]{
\includegraphics[width=.18\textwidth]{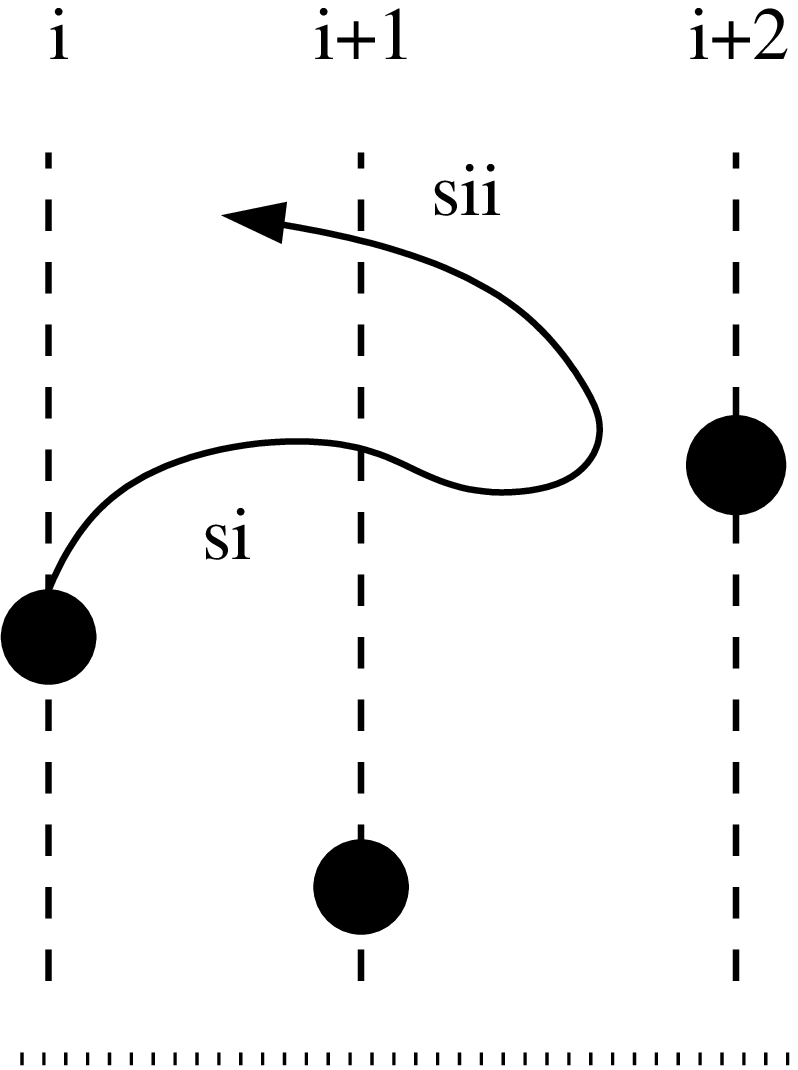}
\label{fig:crossingb}
}%
\caption{Detecting crossings: (a) Two possible particle paths that are
  associated with different braid group generators; (b) Two crossings that
  yield no net braiding.  The reference line used to detect crossings is shown
  dotted, and the perpendicular lines used to determine the braid generator
  are shown dashed.}
\label{fig:crossing}
\end{figure}

Assuming a crossing has occurred between the $\ip$th and $(\ip+1)$th particles,
we need to determine if the corresponding braid generator is~$\sigma_\ip$
or~$\sigma_\ip^{-1}$.  Look at the projection of the $\ip$th and~$(\ip+1)$th
particles in the direction perpendicular to the reference line (the vertical
axis in our case).  If the $\ip$th particle is \emph{above} the~$(\ip+1)$th at
the time of crossing, then the interchange involves the group
generator~$\sigma_\ip$ (we define ``above'' as having a greater value of
projection along the perpendicular direction).  Conversely, if the $\ip$th
particle is \emph{below} the~$(\ip+1)$th at the time of crossing, then the
interchange involves the group generator~$\sigma_\ip^{-1}$.
Figure~\ref{fig:crossinga} depicts these two situations.

The method just described might seem to detect spurious braids if two
well-separated particles just happen to interchange position several times in
a row on the reference line, as shown in Figure~\ref{fig:crossingb}.  However,
this would imply a sequence of~$\sigma_\ip$ and~$\sigma_\ip^{-1}$ braids,
since which particle is the~$\ip$th one changes at each crossing.  When
composed together these crossings produce no net braiding at all.

\begin{figure*}
  \centering
  \psfrag{t}{\small \raisebox{-.5em}{$\time$}}
  \psfrag{eig}{\hspace{-3em}\small
    \raisebox{.7em}{Braiding Factor}}
\subfigure[]{
    \includegraphics[width=.9\columnwidth]{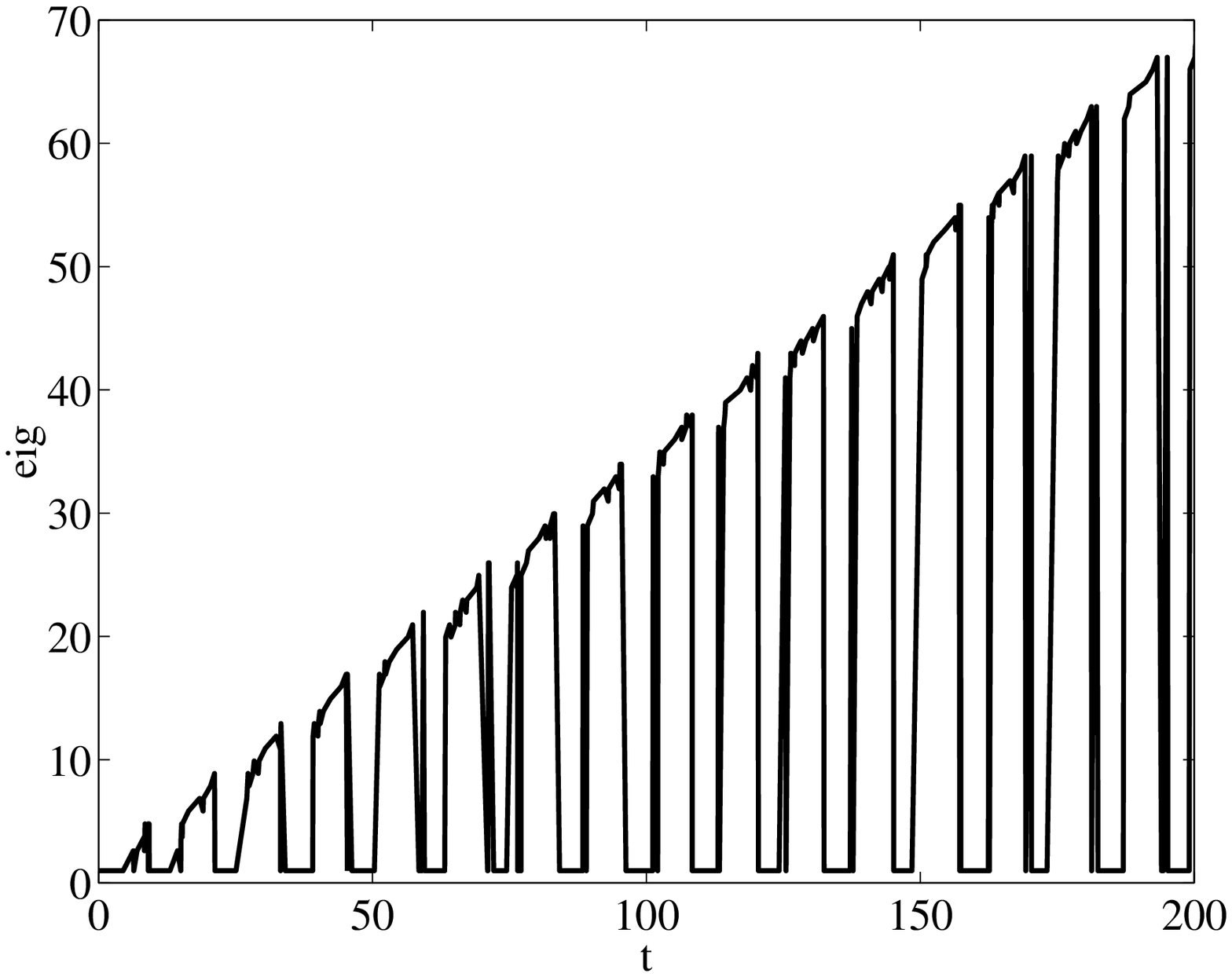}
    \label{fig:eigM_Gamma=0.5}
}\goodgap
\subfigure[]{
    \includegraphics[width=.9\columnwidth]{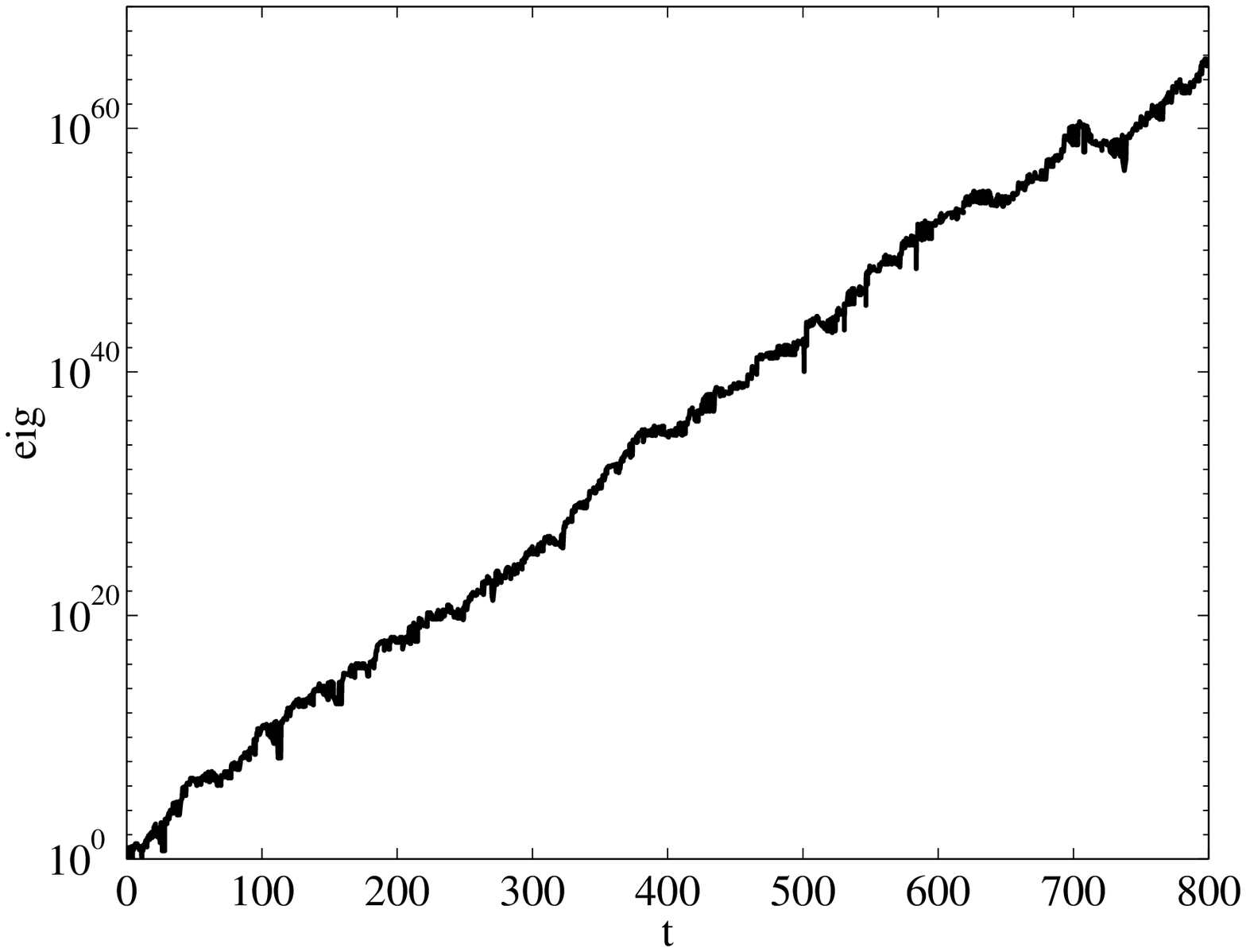}
    \label{fig:eigM_Gamma=13}
}
\caption{Braiding factor (largest eigenvalue of~$\ssigma^{(N)}$) as a function
  of time for a triplet of particles (a) without topological chaos
  ($\Gamma=0.5$); (b) with topological chaos ($\Gamma=13$).  Note that in (b)
  the vertical axis is logarithmic.}
\end{figure*}

We now select a matrix representation for the generators of the braid
group~\cite{Boyland2000,Vikhansky2003}.  These matrices are given by the Burau
representation~\cite{Burau1936,Jones1987} of
% http://mathworld.wolfram.com/BurauRepresentation.html
the~$\nn$-braid group, which consists of~$(\nn-1)\times(\nn-1)$ matrices
defined by
\begin{equation}
  {[\sigma_\ip]}_{k\ell} = \delta_{k\ell} + \delta_{k,\ip-1}\,\delta_{\ell\ip}
  - \delta_{k,\ip+1}\,\delta_{\ell\ip}\,,
  \label{eq:buraurep}
\end{equation}
% More general Burau representation:
%\begin{equation}
%  {[\sigma_\ip]}_{k\ell} = \delta_{k\ell}
%  - \tau\,\delta_{k,\ip-1}\,\delta_{\ell\ip}
%  - \delta_{k,\ip+1}\,\delta_{\ell\ip}
%  - (1+\tau)\,\delta_{k\ip}\,\delta_{\ell\ip},
%\end{equation}
with inverses
\begin{equation}
  {[\sigma_\ip^{-1}]}_{k\ell} =
  \delta_{k\ell} - \delta_{k,\ip-1}\,\delta_{\ell\ip}
  + \delta_{k,\ip+1}\,\delta_{\ell\ip}\,,
  \label{eq:buraurepi}
\end{equation}
%\begin{equation}
%  {[\sigma_\ip^{-1}]}_{k\ell} = \delta_{k\ell}
%  - \delta_{k,\ip-1}\,\delta_{\ell\ip}
%  - \frac{1}{\tau}\,\delta_{k,\ip+1}\,\delta_{\ell\ip}
%  - \Bigl(1+\frac{1}{\tau}\Bigr)\delta_{k\ip}\,\delta_{\ell\ip},
%\end{equation}
where~\hbox{$\ip,k,\ell=1,\ldots,\nn-1$} and we set~$\delta_{k,0}$
and~$\delta_{k,\nn}$ to zero.  (For simplicity, we do not distinguish between
the elements of the braid group and their matrix representation.)  The
determinant of each of these matrices is unity, and they satisfy the
``physical braid'' conditions~\cite{Murasugi}: \hbox{$\sigma_\ip\sigma_j =
\sigma_j\sigma_\ip$} for~\hbox{$|\ip-j|\ge 2$}, and
\hbox{$\sigma_\ip\sigma_{\ip+1}\sigma_\ip =
\sigma_{\ip+1}\sigma_\ip\sigma_{\ip+1}$}.  The
matrices~\eqref{eq:buraurep}--\eqref{eq:buraurepi} can be understood as
arising from the lengthening of line segments tied to the particles as the
particles braid around each other~\cite{Boyland2000,Vikhansky2003}.

As we detect crossings, we compute the running product~$\ssigma^{(N)}$ of all
the braid group elements,
\begin{equation}
  \ssigma^{(N)} = \sigma^{(N)}\cdots\sigma^{(2)}\sigma^{(1)}
\end{equation}
where~\hbox{$\sigma^{(\mu)} \in\{\sigma_\ip\,,\,\sigma_\ip^{-1}\,\lvert\, \ip
= 1,2,\ldots,\nn-1\}$} and~$N(\time)$ is the number of crossings detected
after a time~$\time$.  Now~$\ssigma^{(N)}$ is the product of a sequence of
(possibly random) matrices.  We define the \emph{braiding factor} to be the
largest eigenvalue of~$\ssigma^{(N)}$.  According to Oseledec's multiplicative
theorem~\cite{Oseledec1968}, we can express the time-asymptotic exponential
growth rate of the braiding factor by a (nonnegative) Lyapunov exponent, which
we call the braiding exponent, defined by
\begin{equation}
  \text{braiding exponent} = \lim_{\time\rightarrow\infty}
  \frac{1}{\time}\,\log\,\lvert\,\text{braiding factor}\,\rvert.
\end{equation}
The braiding exponent is a function of the number~$\nn$ of braiding particles.
If the exponent is positive, then we say that the sequence of braids
exhibits~\emph{topological chaos}.  Note that the braiding exponent has units
of inverse time, so that if the frequency of crossings decreases then the
exponent also decreases.

We illustrate the method using the blinking vortex flow (eggbeater flow) of
Aref~\cite{Aref1984}.  The flow consists of two spatially-fixed vortices with
opposite circulation that act alternately in time.  It has the advantage of
being confined to a bounded circular domain and of exhibiting chaos for a
large enough value of the circulation~$\Gamma$.

Figure~\ref{fig:eigM_Gamma=0.5} shows the braiding factor for three particles
($\nn=3$) as a function of time for~$\Gamma=0.5$.  For the initial condition
chosen, only one particle is undergoing chaotic motion; the other two move in
periodic orbits with incommensurate frequencies.  Two features are evident in
Figure~\ref{fig:eigM_Gamma=0.5}: the braiding factor grows more or less
linearly, and it regularly returns to unity.  Because of the linear growth,
the braiding exponent for this triplet of trajectories is zero.

Figure~\ref{fig:eigM_Gamma=13} shows the same system (with the same initial
condition for the triplet of trajectories) but for~$\Gamma=13$, making the
system chaotic almost everywhere.  In this case, the braiding factor grows
exponentially, and it never returns to unity (except at the very beginning).
This is the signature of topological chaos: the braiding exponent for the
triplet is about~$0.2$.  For large enough time, the exponent is the same for
any triplet initially within the same chaotic region.

\begin{figure}
\psfrag{topolyap}{\hspace{-2.9em}\small
  \raisebox{-.75em}{Braiding exponent}}
\psfrag{lyap}{\hspace{-3.75em}\small
  \raisebox{.7em}{Lyapunov exponent}}
%\psfrag{Lyapunov exponent       }{\tiny Lyapunov exponent}
%\psfrag{Line-stretching exponent}{\tiny Line-stretching exponent}
\begin{center}
\includegraphics[width=.9\columnwidth]{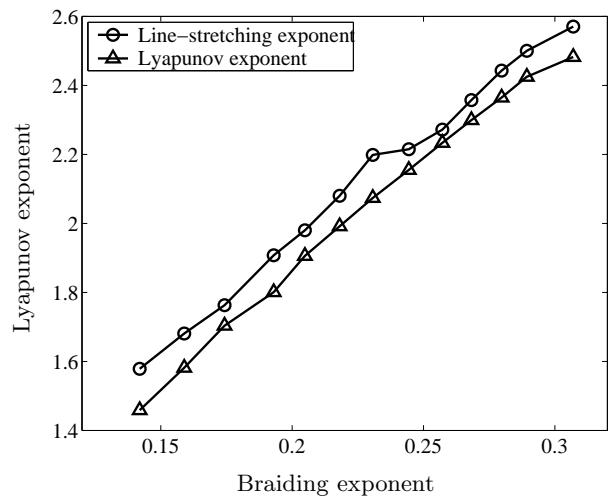}
\end{center}
\caption{Line-stretching and Lyapunov exponents plotted versus the braiding
  exponent for three particles ($\nn=3$) and circulation~$\Gamma$ varying
  from~$8$ to~$20$.  All the exponents increase monotonically with
  circulation.}
\label{fig:lyap_vs_topolyap}
\end{figure}

Of course, the braiding exponent is most useful if it correlates well with the
Lyapunov exponent based on the exponential separation of trajectories.  The
braiding exponent is a lower bound on the rate of stretching of material
lines~\cite{Boyland2000,Vikhansky2003} (the topological entropy), which in
turn is correlated with the Lyapunov exponent.
Figure~\ref{fig:lyap_vs_topolyap} shows the two types of exponents (braiding
and Lyapunov) plotted against each other for different values of the
circulation in the blinking vortex flow.
The relationship between the two is almost linear: both the Lyapunov and
braiding exponents increase monotonically with circulation.  This is
consistent with the exponents measuring the same underlying ``degree of
chaos''.  There is a similar correlation between the braiding exponent and the
rate of stretching of material lines.

\begin{figure}
\psfrag{topolyap}{\hspace{-2.9em}\small
  \raisebox{.75em}{Braiding exponent}}
\psfrag{N}{\hspace{-0em}\small
  \raisebox{-.7em}{$\nn$}}
\psfrag{G = 16.5}{$\Gamma=16.5$}
\psfrag{Odd }{odd}
\psfrag{Even}{even}
\begin{center}
\includegraphics[width=.9\columnwidth]{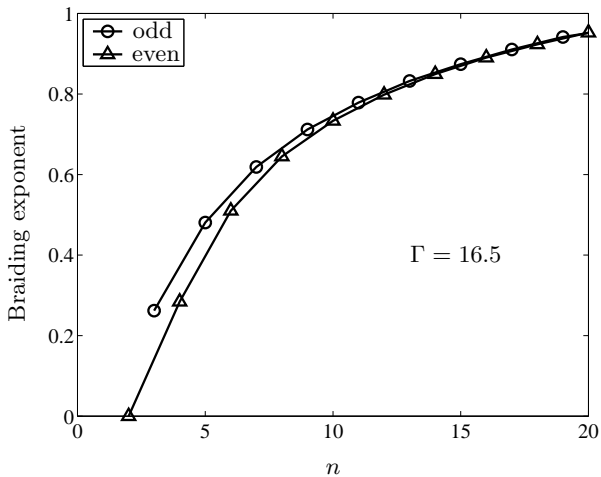}
\end{center}
\caption{Braiding exponent as a function of the number of braiding particle
  orbits, for an odd and even number of particles.}
\label{fig:topolyap_vs_N}
\end{figure}

Figure~\ref{fig:topolyap_vs_N} shows the behavior of the braiding
exponent as a function of the number~$\nn$ of braiding particles.  For~$\nn=1$
the braiding exponent is undefined, since one particle cannot braid around
itself.  For~$\nn=2$ the exponent is always zero (our representation of the
$2$-braid group contains only the identity).  For~\hbox{$\nn\ge 3$} the
exponent evolves along different curves depending on whether~$\nn$ is odd or
even; these converge together for large~$\nn$.  Since the braiding exponent is
a lower bound on the topological entropy of the flow, the curve should
saturate for large~$\nn$: simulations were performed all the way to~$\nn=100$
(not shown) with slow growth of the braiding exponent but no sign of
saturation.  One reason for the growth of the braiding exponent with~$\nn$ is
simply that more braiding particles give more crossings for a given length of
time.  It may thus be advantageous to use many particles when the length of
the time series is limited, as long as the particles exhibit sufficiently
nontrivial braiding (not simply crossing over each other as in
Fig.~\ref{fig:crossingb}).  Note that the $\nn$-particle braiding exponent
exhibits a linear relationship with the Lyapunov exponent, just as it did
for~$\nn=3$ (Fig.~\ref{fig:lyap_vs_topolyap}).

The diagnostic approach presented here sheds some light on Vikhansky's
results~\cite{Vikhansky2003}: he observed an increase in the chaotic
stretching properties of his cavity flow of about~$30\%$ when freely-moving
rods were present.  Compared to the topological effect observed by
Boyland~\etal~\cite{Boyland2000}, this is a modest increase.  This is because
the rods in Vikhansky's case are no more topological obstacles than any fluid
particle.  Their presence modifies the flow and so increases its chaotic
properties, but this is not a topological effect.  In other words, in
Vikhansky's case the chaos is already present even without rods, whereas in
Boyland~\etal\ the rods \emph{cause} the chaos.

The measure of chaos presented here is attractive because of its emphasis of
\emph{global} aspects of the flow.  Lyapunov exponents are difficult to
measure because we try to infer from them a global quantity (the Lyapunov
exponent associated with a given region of the flow) from local measurements
(the rate of separation of nearby trajectories).  These local measurements are
difficult for an experimentalist to make, so we suggest focusing on the global
braiding of particles instead.

\begin{acknowledgments}
The author thanks Francesco Paparella, Joseph Lacasce, Edward Spiegel, Philip
Boyland, Matthew Finn, and Andrew Gilbert for helpful discussions.
\end{acknowledgments}

%\bibliography{books,journals_abbrev,chaotic_advection,braid}

\end{document}